\documentclass[12pt]{article}
\usepackage{latexsym}
\usepackage{epsfig,amssymb,euscript,slashed}
\usepackage{amsmath}
\usepackage[nosort]{cite} 
\usepackage{array,calc,epsfig}
\usepackage{color}
\usepackage{bbm}
\usepackage{fancybox}
\usepackage[linktocpage]{hyperref}

\definecolor{cardinal}{rgb}{0.6,0,0}
\definecolor{darkgreen}{rgb}{0,0.4,0}

\numberwithin{equation}{section} 
\usepackage[]{graphicx}

\begin{document}
\font\cmss=cmss10 \font\cmsss=cmss10 at 7pt

\hfill
\vspace{18pt}
\begin{center}
{\Large 
\textbf{Non-extremal superdescendants of the D1D5 CFT}
}
\end{center}

\vspace{8pt}
\begin{center}
{\textsl{Alessandro Bombini and Stefano Giusto}}

\vspace{1cm}

\textit{\small  Dipartimento di Fisica ed Astronomia ``Galileo Galilei",  Universit\`a di Padova,\\Via Marzolo 8, 35131 Padova, Italy} \\  \vspace{6pt}

\textit{\small  I.N.F.N. Sezione di Padova,
Via Marzolo 8, 35131 Padova, Italy}\\
\vspace{6pt}

\end{center}

\vspace{12pt}

\begin{center}
\textbf{Abstract}
\end{center}

\vspace{4pt} {\small
\noindent 
We construct solutions of IIB supergravity dual to non-supersymmetric states of the D1D5 system. These solutions are constructed as perturbations carrying both left and right moving momentum around the maximally rotating D1D5 ground state at linear order. They are found by extending to the asymptotically flat region the geometry generated in the decoupling limit by the action of left and right R-currents on a known D1D5 microstate. The perturbations are regular everywhere and do not carry any global charge. We also study the near-extremal limit of the solutions and derive the first non-trivial correction to the extremal geometry.}

\vspace{1cm}


\thispagestyle{empty}

\vfill
\vskip 5.mm
\hrule width 5.cm
\vskip 2.mm
{
\noindent  {\scriptsize e-mails:  {\tt alessandro.bombini@pd.infn.it, stefano.giusto@pd.infn.it} }
}

\setcounter{footnote}{0}
\setcounter{page}{0}

\newpage

\section{Introduction}
The D1D5 system has been a useful physical set-up to study the microscopic properties of black holes within String Theory since the very early days \cite{Strominger:1996sh}. When the $S^1$ common to the $n_1$ D1 and the $n_5$ D5 branes is taken larger than all other scales in the problem, the system is described by a CFT and holographic methods can be applied to gain insight on the system. The CFT becomes particularly simple at a special point of the theory's moduli space where it reduces to an orbifold sigma-model with target space $\mathcal{M}^N/S^N$, with $\mathcal{M}$ the compact space wrapped by the D5's and $N=n_1 n_5$. The orbifold CFT provides useful informations on many aspects of black hole physics: it can be used for example for counting extremal states that carry the minimum allowed energy for given charges \cite{Dijkgraaf:1996xw,Maldacena:1999bp}, or for computing moduli-independent quantities, like 3-point functions of 1/4 and 1/8 BPS operators \cite{Kanitscheider:2006zf,Kanitscheider:2007wq,Giusto:2015dfa}. Sometimes the orbifold CFT gives exact results even for processes that are not expected to be protected by supersymmetry, like for travel times and decay rates of perturbations in supersymmetric \cite{Lunin:2001dt} and non-supersymmetric states \cite{Avery:2009tu}. However many qualitative features of black holes, like thermalisation \cite{Balasubramanian:2011ur} or chaos \cite{Maldacena:2015waa}, are not captured by the orbifold CFT, and can be analysed only by deforming the CFT away from the free orbifold point. This can be done perturbatively by turning on marginal twist operators \cite{Avery:2010er, Carson:2014ena, Burrington:2014yia}, but extrapolating this perturbative expansion all the way to the strong coupling regime where the supergravity description applies, seems a daunting task.  

At the supergravity point a simple picture of black hole microstates is expected again to emerge: the CFT states that can be identified with the microstates of black holes have a conformal dimension that scales like the central charge $c=6 N$ and are expected to be described by smooth horizonless geometries with the same global charges as the black hole \cite{Mathur:2005zp,Skenderis:2008qn, Balasubramanian:2008da,Chowdhury:2010ct,Bena:2013dka}. In the ``decoupling limit" in which the holographic description applies, these ``microstate geometries" tend at large distances to AdS$_3\times S^3\times \mathcal{M}$. Establishing a precise link between the geometries and the states of the orbifold CFT is however a subtle task, already for supersymmetric states. One source of complications is that not all states that preserve some supersymmetry at the orbifold point are guaranteed to exist at strong coupling: as the moduli are changed, short multiplets might join into long multiplets whose conformal dimension grows without bound in the supergravity limit. This does not happen for 1/4 BPS states, and indeed one can construct all the geometries dual to Ramond-Ramond ground states \cite{Lunin:2001jy,Kanitscheider:2007wq} and establish a precise map with the orbifold states \cite{Kanitscheider:2006zf}. However with this much supersymmetry there is no supergravity solution with a regular horizon of finite area, and these states represent the microstates of string-size black holes. For 1/8 BPS configurations, for which a regular black hole exists already at the classical level, the situation is more involved. A subset of states that can be matched with supergravity modes has been identified in \cite{deBoer:1998ip}, and is dubbed ``supergraviton gas". The characterisation of the supergraviton gas relies on the CFT chiral algebra, which consists of the usual left and right Virasoro generators ($L_n, \tilde L_n$), of an affine $SU(2)_L\times SU(2)_R$ R-symmetry algebra ($J^a_n, \tilde J^a_n$), corresponding to rotations in the non-compact spatial dimensions, and of the $(4,4)$ supercharges ($G^{\pm,\pm}_n, \tilde G^{\pm,\pm}_n$). One should also recall that the states of an orbifold theory split in different twist sectors: each twist sector can be described as a collection of effective strings, or ``strands'', of different winding numbers, with the constraint that the total winding of all the strands equals $N$. Graviton gas states are the ones obtained by acting globally on a 1/4 BPS state with any element of the CFT chiral algebra (we will refer to such states as ``superdescendants") or by acting independently on any strand with an element of the global sub-algebra generated by $L_0$, $L_\pm 1$, $J^a_0$, $G^{\pm,\pm}_{\pm 1/2}$ plus the corresponding right-moving generators. The geometries dual to this latter class of states -- known as ``superstrata" -- have been constructed in \cite{Bena:2015bea,Bena:2016agb,Bena:2016ypk} exploiting a crucial linear structure possessed by the BPS equations \cite{Bena:2011dd}: one constructs the building blocks corresponding to each individual strand and then uses the linearity of the equations to take linear combinations of these building blocks and thus generating the geometry of the full state. 

Generating the geometries of superdescendants is much simpler. Indeed the CFT chiral algebra is represented on the gravity side by diffeomorphisms that do not vanish at the AdS boundary and the geometries of superdescendants are obtained by applying such diffeomorphisms to the RR ground state geometries. This produces asymptotically AdS solutions: black holes, however, are asymptotically flat (by which we mean that at large distances the spacetime is $\mathbb{R}^{4,1}\times S^1\times \mathcal{M}$), and geometries that admit an interpretation as black hole microstates should be extendable from asymptotically AdS solutions to asymptotically flat ones. Even for supersymmetric superdescendants this extension requires solving a non-trivial gravity problem, since the asymptotically flat geometry is not diffeomorphic to the seed 1/4 BPS solution, with the non-trivial deformation of the geometry occurring in the ``neck'' region that joins AdS and asymptotic infinity. This problem was first considered in \cite{Mathur:2003hj}, where it was solved using a double approximation: first one considers the limit in which the microstate can be described as a perturbation around the background of a simpler state, then the linear equations for the perturbation are solved approximately using a matching procedure between the AdS and the asymptotic regions. This technique was generalised in \cite{Mathur:2011gz,Mathur:2012tj,Shigemori:2013lta}, while an exact construction was given, for two different classes of states, in \cite{Lunin:2012gp} and \cite{Giusto:2013bda}. The existence of an exact solution is again ultimately a consequence of the linearity of the BPS equations.

The purpose of this paper is to investigate how much of this structure extends to non-supersymmetric microstates. We know only very few non-extremal microstate geometries. The first example, and also the only one with a known CFT dual, was constructed in \cite{Jejjala:2005yu} generalising to the non-BPS case the technique of \cite{Giusto:2004ip}. The full holographic interpretation of this solution was found in \cite{Chakrabarty:2015foa} and it involves states obtained by applying spectral flow in both left-moving and right-moving sectors to some simple 1/4 BPS state. The existence of non-supersymmetric supergravity solutions that carry no global charges was conjectured in \cite{Mathur:2013nja}, where a construction of these solutions based on neutral oscillating supertubes was outlined. Powerful techniques to construct exact fully non-linear non-supersymmetric solutions have been developed over the past years \cite{Bossard:2014yta,Bossard:2014ola,Bena:2015drs,Bena:2016dbw}, but the relation between these gravity solutions and the states of the CFT is unclear yet. In fact, the issue of which states of the orbifold theory survive in the spectrum at strong coupling is even less understood for non-supersymmetric states than for supersymmetric ones. The existence of a non-BPS analogue of the graviton gas made of states that are not descendants of RR ground states is far from obvious, since the linear properties of the gravity equations that allowed the construction of the supergraviton gas in the supersymmetric setting is not guaranteed to persist when supersymmetry is broken. On the other hand the chiral algebra guarantees the existence of  superdescendants on the whole moduli space, even when the states contain both left-moving and right-moving generators and, hence, break supersymmetry completely. The geometry of these states in the decoupling limit is obtained straightforwardly by acting with a large diffeomorphism on a 1/4 BPS solution. The non-trivial task is again the extension of the asymptotically AdS geometry so constructed to an asymptotically flat one, which could be interpreted as a black hole microstate. We will not attempt here to perform this task at the full non-linear level, and we will work in the regime in which the microstate is a linear perturbation around a supersymmetric background. We show that the simplification that allowed for a simple solution of the problem for supersymmetric states does not happen when the perturbation breaks supersymmetry. We reduce the problem to the solution of a partial differential equation, which we solve approximately using the same matching technique of \cite{Mathur:2003hj}. 

Our main goal is to prove the existence of a solution that interpolates between the geometry in the decoupling limit and a well-behaved solution at large distances. The existence of such a solution is not obvious since non-extremality has drastic effects on the large-distance behaviour of the geometry. It was shown in \cite{Mathur:2003hj} that when the perturbation carries more energy than its charge, it will be non-normalisable, i.e. it belongs to the continuum spectrum of excitations around the extremal background: this behaviour is expected for non-extremal perturbations, and does not signal a pathology of the solution\footnote{We thank Samir Mathur for clarifying to us this point.}. A similar conclusion was reached more recently in \cite{Chakrabarty:2015foa}, which also considered non-extremal states obtained by applying left and right-moving chiral algebra transformations to 1/ 4 BPS geometries: it was concluded that the only normalisable solutions are given by {\it extremal} perturbations around the non-extremal background of \cite{Jejjala:2005yu}. Here we consider genuinely non-extremal perturbations and we find that they indeed fall-off very slowly (like $r^{-3/2}$) at large distances. This is the expected asymptotic behaviour for non-extremal states; moreover we verify that, via a somewhat non-trivial mechanism, the perturbation does not alter the global charges of the background. Hence we conclude that the perturbative non-extremal solutions we find can be consistently identified with non-supersymmetric microstates of the black hole. 

In the next Section we describe the states we consider using the language of the orbifold CFT, and also recall the construction of their dual geometries in the decoupling limit. We introduce the general ansatz we use to extend the asymptotically AdS geometries to the asymptotically flat region in Section \ref{sec:ansatz}: we consider a background dual to the maximally rotating RR ground state and a perturbation that excites the NSNS B-field and the RR 0-form and write down an ansatz that generalises the one appropriate for supersymmetric solutions \cite{Giusto:2013rxa}. In Section \ref{sec:equations} we work out the linearised equations of motion for the perturbation and we reorganise them in such a way that they reduce to a single partial differential equation of the second order. The whole perturbation can be easily reconstructed from a solution of this differential equation. We solve approximately this differential equation by matching at leading order the decoupling limit geometry, valid at small distances, with the solution at large distances. We do this for two different classes of states. In Section \ref{sec:nonextremal} we consider a maximally non-extremal state obtained by adding to the maximally rotating RR ground state an equal amount of left and right-moving momentum via an $SU(2)_L\times SU(2)_R$ affine algebra transformation. In Section \ref{sec:nearextremal} we construct a near-extremal state, where we take the left-moving momentum much bigger than the right-moving one and we solve the equations at the first non-trivial order in the non-extremality parameter. We summarise our results and outline some possible developments in the concluding Section. The details of some calculations are collected in the appendices: in Appendix \ref{sec:appeqs} we adapt the equations of motion to our ansatz and simplify them, in Appendix \ref{sec:lapl} we prove a useful identity for the covariant Laplacian.  

\section{The CFT states}
\label{sec:CFT}

A system of $n_1$ D1 branes and $n_5$ D5 branes compactified on $S^1\times \mathcal{M}$ is described, in the limit in which the radius $R$ of the $S^1$ is much larger than both the string scale and the volume of $\mathcal{M}$, by a 2-dimensional CFT with $(4,4)$ supersymmetry. The CFT admits a simple description at a special point of its moduli space, where it reduces to a sigma-model on the orbifold space $\mathcal{M}^N/S^N$ with $N=n_1 n_5$. We refer to \cite{Avery:2010qw} and \cite{Giusto:2015dfa} for an introduction on the orbifold theory and for the details on our conventions. The basic ingredients that will be useful for us here are the RR ground states of the theory and its symmetry algebra.   

The RR ground states can be described as a collection of ``strands'', each characterised by the winding number $k$ and the left/right R-charge quantum numbers $(\jmath,\bar \jmath)$: we will denote the state of each strand by $|\jmath,\bar \jmath\rangle_k$ and the full D1D5 state containing $N_i$ strands of type $|\jmath_i,\bar \jmath_i\rangle_{k_i}$ as $\prod_i (|\jmath_i,\bar \jmath_i\rangle_{k_i})^{N_i}$, with the constraint that the total winding number equals $N$, i.e. $\sum_i k_i N_i=N$. For example the RR ground state with the maximum values of $(\jmath,\bar \jmath)$ is denoted as $(|+,+\rangle_1)^N$ and has $\jmath=\bar \jmath=N/2$. The CFT has a spectral flow symmetry which maps R and NS sectors. Performing one unit of spectral flow on the left and the right sector of the CFT maps the state $(|+,+\rangle_1)^N$ into the SL(2,$\mathbb{C}$)-invariant vacuum. This fact allows to deduce easily the gravity dual geometry for the state $(|+,+\rangle_1)^N$.

Each D1D5 ground state admits a dual description in terms of an asymptotically AdS$_3\times S^3\times \mathcal{M}$ geometry \cite{Lunin:2001jy}. For all the states considered in this work the compact 4D space $\mathcal{M}$ will just play a spectator role, and we will only focus on the dimensionally reduced 6-dimensional theory. The geometry dual to the SL(2,$\mathbb{C}$)-invariant vacuum is simply global AdS$_3\times S^3$:
\begin{subequations}\label{eq:6Dvacuum}
\begin{align}
ds^2&=\sqrt{Q_1 Q_5}\,(ds^2_{AdS_3}+ds^2_{S^3})\,,\\
ds^2_{AdS_3}&=\frac{d r^2}{r^2+a^2}-\frac{r^2+a^2}{Q_1Q_5}dt^2+\frac{r^2}{Q_1Q_5}dy^2\,,\label{eq:AdSmetric}\\
ds^2_{S^3}&=d \theta^2+\sin^2\theta\, d\hat{\phi}^2+\cos^2\theta\, d\hat{\psi}^2\,,\label{eq:Smetric}\\
F&=2 Q_5\,(-\mathrm{vol}_{AdS_3}+\mathrm{vol}_{S^3})\,,\quad e^{2\Phi}=\frac{Q_1}{Q_5}\,,\\
\mathrm{vol}_{AdS_3}&=\frac{r}{Q_1Q_5}\,d r\wedge dt \wedge dy\,,\quad \mathrm{vol}_{S^3}=\sin\theta \cos\theta\, d \theta\wedge  d\hat{\phi} \wedge d\hat{\psi}\,,
\end{align}
\end{subequations}
where $ds^2$ is the Einstein metric in 6D, $F$ is the RR 3-form field strength and $\Phi$ the dilaton. $Q_1$ and $Q_5$ are the supergravity D1 and D5 charges
\begin{equation}
Q_1 = \frac{(2\pi)^4 n_1 g_s (\alpha')^3}{V_4}\,,\quad Q_5 =  g_s n_5 \alpha'\,,
\end{equation}
where $g_s$ is the string coupling and $V_4$ is the volume of the $\mathcal{M}$. The parameter $a$ is linked to the D-brane charges and the $S^1$ radius $R$ by
\begin{equation}\label{eq:a}
a=\frac{\sqrt{Q_1 Q_5}}{R}\,.
\end{equation}
In the following we will slightly simplify our equations by taking $Q_1=Q_5=Q$.  Spectral flow acts geometrically on the gravity side through the change of coordinates
\begin{equation}\label{eq:sf}
\hat \phi = \phi - \frac{t}{R}\,,\quad \hat \psi = \psi - \frac{y}{R}\,.
\end{equation}
Note that this is a diffeomorphism that acts non trivially at the AdS$_3$ boundary, and hence it changes the state. Thus the geometry dual to the state $(|+,+\rangle_1)^N$ is (\ref{eq:6Dvacuum}) with the coordinate redefinition (\ref{eq:sf}).

One can construct more interesting and more generic states by adding strands with different winding numbers and/or different spins. For example one can consider the state $|0,0\rangle_k$, with winding $k$ and $\jmath=\bar\jmath =0$ (there is only one such state that is also a scalar under the $SU(2)\times SU(2)$ group that corresponds to rotations in the internal space $\mathcal{M}$). The RR ground state $(|+,+\rangle_1)^{N_1} (|0,0\rangle_k)^{N_2}$ with $N_1+k N_2=N$ (or more precisely a coherent superposition \cite{Skenderis:2006ah} of states of these form, see \cite{Giusto:2015dfa} for details) sources a non-trivial geometry when both $N_1$ and $N_2$ are numbers of order $N$. The full geometry is given for example in eq. (3.11) of \cite{Giusto:2013bda}. The limit of interest in this paper is when the strands of type $|0,0\rangle_k$ are much fewer then the ones of type $|+,+\rangle_1$: $N_2 \ll N_1$. In this regime the appropriate gravitational description of the state is as a perturbation around the AdS$_3\times S^3$ background (\ref{eq:6Dvacuum}), which solves the supergravity equations at linear order. After flowing to the NS sector this perturbation is an anti-chiral primary of dimension and charge $h=\bar h=-\jmath=-\bar\jmath = k/2$ \cite{Mathur:2003hj}. The linearised perturbation is controlled by a scalar $w$, which is identified with the RR 0-form, and by the 2-form B-field $B$, and is given by
\begin{subequations}\label{eq:pert2charge}
\begin{align}
w = \mathcal{B}\,Y\,,\quad B=\frac{Q}{k} (Y *_{AdS_3} d\mathcal{B}- \mathcal{B} *_{S^3} dY)\,,~~~~~~~~~~~\\
\mathcal{B}= \frac{b\,R}{Q}\, \left(\frac{a}{\sqrt{r^2+a^2}}\right)^{\!k} e^{-i k \frac{t}{R}}\,,\quad Y = Y^{\ell,\ell}_{-\ell,-\ell} = \sin^k\theta \,e^{-i k \hat \phi}\quad \mathrm{with}\quad k = 2\ell\,,
\end{align}
\end{subequations}
where $*_{AdS_3}$ and $*_{S^3}$ are the Hodge duals\footnote{Our conventions for the Hodge dual of a $p$-form in $d$ dimensions are\begin{equation}
* \omega^{(p)} = \frac{\sqrt{|g|}}{p! (d-p)!}{\epsilon_{i_1\ldots i_{d-p}}}^{j_1\ldots j_p}\omega^{(p)}_{j_1\ldots j_p}dx^{i_1}\wedge \ldots dx^{i_{d-p}}\,.\nonumber
\end{equation}
We choose the 3D orientations so that $\epsilon_{rty}=\epsilon_{\theta\phi\psi}=+1$, the 4D orientation so that $\epsilon_{r\theta\phi\psi}=+1$ and the 6D orientation so that $\epsilon_{rty\theta\phi\psi}=+1$.}  with respect to the AdS$_3$ and $S^3$ metrics defined in (\ref{eq:AdSmetric}) and (\ref{eq:Smetric}). $Y^{\ell,\ell}_{\jmath,\bar \jmath}$ denotes the $S^3$ spherical harmonics of order $k=2\ell$:
\begin{equation}
\Box_{S^3} Y^{\ell,\ell}_{\jmath,\bar \jmath} = - k(k+2) Y^{\ell,\ell}_{\jmath,\bar \jmath}\,;
\end{equation} 
analogously $\mathcal{B}$ is an eigenfunction of the  AdS$_3$  Laplacian:
\begin{equation}
\Box_{AdS_3} \mathcal{B} = k(k-2) \mathcal{B}\,.
\end{equation}
The parameter $b$ controls the number of strands of type $|0,0\rangle_k$ ($b^2 \sim N_2/N$) and the above perturbation solves the supergravity equations at first order in $b$.

While the holographic description of RR ground states is well-understood \cite{Lunin:2001jy,Kanitscheider:2006zf,Kanitscheider:2007wq}, the analysis of excited states, and in particular of non-supersymmetric states which carry excitations on both the left and right sectors of the CFT, is largely incomplete. An effective way to approach the problem is to use the CFT chiral algebra, which, for general $\mathcal{M}$, it is composed by the usual Virasoro generators, by the $SU(2)_L\times SU(2)_R$ R-currents $J^a_{n}$, $\tilde J^a_n$ and by the fermionic supercurrents. A simple class of excited states, which are guaranteed to exist at any point of the CFT moduli space, is formed by descendants obtained by acting on RR ground states with an arbitrary string of chiral algebra generators. In this paper we focus on the R-charge currents and consider the states 
\begin{equation}\label{eq:nonsusystate}
[|+,+\rangle_1]^{N_1} [(J^+_{-1})^m (\tilde J^+_{-1})^{\bar m} \,|0,0\rangle_k]^{N_2}\,;
\end{equation}
when both $m$ and $\bar m$ are non-vanishing theses states are non-supersymmetric. As explained above, in the limit $N_2\ll N_1$ the states are holographically described by a linearised perturbation around AdS$_3\times S^3$, which is easily constructed starting from (\ref{eq:pert2charge}). Indeed, when one flows to the NS sector the currents $J^+_{-1}$ and $\tilde J^+_{-1}$ are mapped to $J^+_{0}$ and $\tilde J^+_{0}$, which rotate the perturbation (\ref{eq:pert2charge}) while leaving the AdS$_3\times S^3$ background invariant. Thus the perturbation dual to the state (\ref{eq:nonsusystate}) for $N_2\ll N_1$ is of the form of eq. (\ref{eq:pert2charge}), with the same $\mathcal{B}$ but a rotated $Y$:
\begin{equation}\label{eq:rotatedY}
Y = Y^{\ell,\ell}_{-\ell+m,-\ell+{\bar m}}\,.
\end{equation}

This construction provides a systematic way to generate the geometries dual to descendant states in the decoupling limit, valid when the $S^1$ radius is large ($R\gg \sqrt{Q}$) and in the inner region of the spacetime, in which $r\ll \sqrt{Q}$. In this region the geometries have AdS$_3\times S^3$ asymptotics. If the states represent microstates of asymptotically flat black holes, they should admit a description outside the inner region, in which they smoothly join to the $\mathbb{R}^{4,1}\times S^1$ flat spacetime at large distances. The construction of this asymptotically flat extension for non-supersymmetric states will be the focus of the remainder of this article. We will concentrate on two subclasses of states: non-extremal states with $m=\bar m=1$ and near-extremal states with $m=k\gg 1$, $\bar m=1$.

\section{Asymptotically flat ansatz}
\label{sec:ansatz}

The non-supersymmetric solutions we are after solve the equations of motion of type IIB supergravity linearised around a supersymmetric background, which is the asymptotically flat extension of the AdS$_3\times S^3$ solution (\ref{eq:6Dvacuum}). We will work in the 6D theory dimensionally reduced on $\mathcal{M}$, whose field content and equations of motion have been nicely reviewed in \cite{Roy:2016zzv}. The fields that make up the background are the metric, the RR 3-form field strength $F=d C$, the dilaton and the volume of $\mathcal{M}$; these last two scalars trivialise if one takes $Q_1=Q_5$. The background, which represents the first example~\cite{Balasubramanian:2000rt,Maldacena:2000dr} of an asymptotically flat solution dual to a D1D5 state (the maximally rotating RR ground state $|+,+\rangle_1^{N}$), can be conveniently written as
\begin{subequations}\label{eq:background}
\begin{align}
ds^2=&-\frac{2}{Z}(du+\omega)(dv+\beta)+Z\,ds^2_4\,,\\
C=&-\frac{1}{Z}(d u +\omega)\wedge (dv+\beta) + \gamma\,,
\end{align}
\end{subequations}
with
\begin{subequations}\label{eq:background2}
\begin{align}
ds^2_4 &= \Sigma \left(\frac{dr^2}{r^2+a^2}+d\theta^2 \right)+(r^2+a^2)\sin^2\theta d\phi^2+r^2\cos^2\theta d\psi^2\,,\\
\beta &= \frac{R\,a^2}{\sqrt{2}\,\Sigma}(\sin^2\theta\,d\phi-\cos^2\theta\,d\psi)\,,\quad \omega = \frac{R\,a^2}{\sqrt{2}\,\Sigma}(\sin^2\theta\,d\phi+\cos^2\theta\,d\psi)\,,\\
Z&=1+\frac{Q}{\Sigma}\,,\quad \gamma= - Q\,\frac{r^2+a^2}{\Sigma}\,\cos^2\theta\,d\phi\wedge d\psi\,,\quad \Sigma\equiv r^2+a^2\cos^2\theta\,.
\end{align}
\end{subequations}
The light-cone coordinates $u$ and $v$ are related with time $t$ and the $S^1$ coordinate $y$ by
\begin{equation}
u = \frac{t-y}{\sqrt{2}}\,,\quad v = \frac{t+y}{\sqrt{2}}\,;
\end{equation}
the Euclidean 4D metric $ds^2_4$ is just flat $\mathbb{R}^4$ written in a convenient system of coordinates, which are related to the usual cartesian coordinates $x_i$ by
\begin{equation}
x_1+i x_2 = \sqrt{r^2+a^2}\sin\theta \,e^{i\phi}\,,\quad x_3+i x_4 = r \cos\theta \,e^{i\psi}\,.
\end{equation}
Note that the following relations, which are ultimately a consequence of supersymmetry, are satisfied:
\begin{equation}
d\beta= *_4 d\beta\,,\quad d\omega = - *_4 d\omega\,,\quad *_4 d Z = d\gamma\,,
\end{equation}
where $*_4$ is the Hodge dual with respect to $ds^2_4$. The length scale $a$ is defined in (\ref{eq:a}). It is simple to verify that the asymptotically flat geometry (\ref{eq:background},\ref{eq:background2}) reduces to the AdS$_3\times S^3$ solution (\ref{eq:6Dvacuum}) in the decoupling limit ($r,a\ll \sqrt{Q}$), in which one can discard the 1 in the function $Z$. Note also that for $Q_1=Q_5$ the 3-form $F$ is anti-self-dual in the full asymptotically flat geometry:
\begin{equation}\label{eq:asdF}
F + * F=0\,,
\end{equation}
where $*$ denotes the Hodge dual with respect to the 6D Einstein metric $ds^2$.

The perturbation that corresponds to add few strands of the type $(J^+_{-1})^m (\tilde J^+_{-1})^{\bar m} \,|0,0\rangle_k$ excites the NSNS B-field $B$, the RR 0-form $\chi_1$ and the component of the RR 4-form along $\mathcal{M}$, $\chi_2$: again a slight simplification happens for $Q_1=Q_5$, in which case $\chi_1=\chi_2\equiv w$. In the decoupling limit the form of the perturbation is given by (\ref{eq:pert2charge}) and (\ref{eq:rotatedY}). We find that the task of extending the perturbation to the asymptotically flat region is simplified by using an ansatz inspired by the supersymmetric solutions:
\begin{equation}\label{eq:pertgen}
w= \frac{Z_4}{Z}\,,\quad B = - \frac{Z_4}{Z^2}\,(du+\omega)\wedge (dv+\beta) + a_4\wedge (dv+\omega)+b_4\wedge (du+\beta) + \delta_2\,.
\end{equation}
Here $Z$, $\beta$ and $\omega$ are the same 0 and 1-forms that appear in the background (\ref{eq:background}), while the 0-form $Z_4$, the 1-forms $a_4$, $b_4$ and the 2-form $\delta_2$ are the unknowns that parametrise the perturbation. All these forms have legs only along the 4D Euclidean space $ds^2_4$, but might depend also on $u$ and $v$. It was found in \cite{Giusto:2013rxa} that general supersymmetric solutions have the form (\ref{eq:pertgen}) with $b_4=0$ -- if one specialises the results of \cite{Giusto:2013rxa} to $Q_1=Q_5$; moreover supersymmetry implies that nothing can depend on $u$. It is clear that any 0-form $w$ and any 2-form $B$ can be written as in (\ref{eq:pertgen}), for some choice of $Z_4$, $a_4$, $b_4$, $\delta_2$; having chosen the $uv$ component of $B$ to be controlled by the same function $Z_4$ that appears in $w$ has partially restricted the 2-form gauge invariance $B\to B + d\lambda$. The remaining gauge freedom is the one where $\lambda$ is a $u$ and $v$-dependent 1-form with only legs on $\mathbb{R}^4$; it acts on our unknowns as
\begin{equation}
a_4\to a_4 - \partial_v \lambda\,,\quad b_4\to b_4 - \partial_u \lambda \,,\quad \delta_2\to \delta_2 + \mathcal{D} \lambda\,,
\end{equation}
where we introduce the covariant differential
\begin{equation}
\mathcal{D} \equiv d_4 - \beta \wedge \partial_v - \omega \wedge \partial_u\,,
\end{equation}
with $d_4$ the exterior differential with respect to the $\mathbb{R}^4$ coordinates. The combinations of $a_4$, $b_4$, $\delta_2$ that are left invariant by this residual gauge freedom are
\begin{equation}\label{eq:definitions}
\mathcal{A}\equiv \partial_u a_4 - \partial_v b_4\,,\quad \Theta_4 \equiv \mathcal{D} a_4 + \partial_v \delta_2\,,\quad  \tilde\Theta_4 \equiv \mathcal{D} b_4 + \partial_u \delta_2\,,\quad \Xi\equiv  \mathcal{D} \delta_2 -a_4 \wedge d \beta -b_4 \wedge d \omega \,,
\end{equation}
and it will be convenient to express the equations of motion in terms of these gauge-invariant quantities. These quantities satisfy the Bianchi identities
\begin{subequations}
\begin{align}
\partial_u \Theta_4-\partial_v \tilde \Theta_4= \mathcal{D}\mathcal{A}\,,~~~~~~~~~~~~~~~~~~~~\label{eq:bianchi1}\\
 \mathcal{D} \Theta_4-\partial_v \Xi +\mathcal{A}\wedge d\omega=0\,,\quad  \mathcal{D} \tilde\Theta_4-\partial_u \Xi -\mathcal{A}\wedge d\beta=0\,,\label{eq:bianchi2}\\
 \mathcal{D}\, \Xi = -\Theta_4\wedge d\beta - \tilde \Theta_4\wedge d\omega\,.~~~~~~~~~~~~~\label{eq:bianchi3}
\end{align}
\end{subequations}

For supersymmetric solutions, for which $b_4=0$ and $\partial_u$ is an isometry, $\mathcal{A}$ and $\tilde \Theta_4$ are trivial. In that case the parametrisation (\ref{eq:pertgen}) was particularly useful as it trivialised the problem of finding the asymptotically flat linearised solution, given the one in the decoupling limit \cite{Giusto:2013rxa}. It turns out, indeed, that when supersymmetry is preserved the supergravity equations for $Z_4$, $a_4$ and $\delta_2$ do not depend on $Z$: then, all one has to do to construct the asymptotically flat solution is to keep the same $Z_4$, $a_4$ and $\delta_2$ of the inner region solution and simply ``add back the 1" in the function $Z$ that appears in \cite{Giusto:2013rxa}. The hope is that a similar simplification also happens for non-extremal solutions: we will see that life is not quite as easy, since the equations of motion couple $\mathcal{A}$ with $Z$ and hence deforming $Z$, as it is required by asymptotic flatness, necessarily induces deformations of all the objects ($Z_4$, $a_4$, $b_4$, $\delta_2$) that control the $(w,B)$ fields. Nevertheless we find that using the parametrisation (\ref{eq:pertgen}) helps in simplifying the equations and ultimately reduces the whole problem to a single partial differential equation for a scalar function.

\section{Linearised supergravity equations}
\label{sec:equations}

Our goal is to construct a solution of the linearised equations of motion around the background (\ref{eq:background},\ref{eq:background2}); the solution contains the fields $w$ and $B$, parametrised as in (\ref{eq:pertgen}), and must reduce to the near-horizon solution described in Section~\ref{sec:CFT} in the inner region.

The non-trivial equations of motion for $(w,B)$ at linear order are
\begin{subequations}\label{eq:motionlinear}
\begin{align}
d(* H +2 w F)=0\,,\label{eq:firstzero}\\
d*d w + F \wedge H =0\,,\label{eq:second}
\end{align}
\end{subequations}
where $H=d B$ is the NSNS 3-form and the Hodge dual $*$ and the 3-form $F=dC$ refer to the background (\ref{eq:background},\ref{eq:background2}). The first equation can be partially integrated to
\begin{equation}\label{eq:first}
*H - H + 2 w F=0\,,
\end{equation}
after taking into account the anti-self-duality of $F$ (\ref{eq:asdF}). With the ansatz (\ref{eq:pertgen}), eq. (\ref{eq:first}) is equivalent to 
\begin{subequations}\label{eq:firstbis}
\begin{align}
*_4 \mathcal{D} Z_4   = \,\, \Xi -  Z^2 *_4 \mathcal{A}\,, ~~~~\label{eq:firstone}\\
 \Theta_4 = *_4 \Theta_4 \,,\quad  \tilde \Theta_4 = - *_4 \tilde \Theta_4 \,.\label{eq:firsttwo}
\end{align}
\end{subequations}
The scalar eq. (\ref{eq:second}) adds one more differential constraint which, after using (\ref{eq:firstone}), can be shown to reduce to
\begin{equation}\label{eq:secondone}
*_4 \mathcal{D} *_4 \mathcal{A} = 2 \,\partial_u\partial_v Z_4\,.
\end{equation}
Details of these manipulations can be found in Appendix~\ref{sec:appeqs}.

One can check that the near-horizon solution (\ref{eq:pert2charge},\ref{eq:rotatedY}), when rewritten in the form (\ref{eq:pertgen}), indeed satisfies eqs. (\ref{eq:firstone}), (\ref{eq:firsttwo}) and (\ref{eq:secondone}). When one considers the same equations in the asymptotically flat background, one has to replace $Z\to Z+1$: then the $Z$-dependent term in eq. (\ref{eq:firstone}) is modified, and this induces a non-trivial modification of all other fields. We have already remarked that this complication is peculiar to non-supersymmetric solutions, for which $\mathcal{A}\not =0$. 

Eqs. (\ref{eq:firstone}), (\ref{eq:firsttwo}), (\ref{eq:secondone}) seem to form a complicated set of coupled partial differential equations. One can however considerably simplify the problem by reducing this set to a single equation for the 1-form $\mathcal{A}$. This is done as follows. Eqs. (\ref{eq:bianchi1}) and (\ref{eq:firsttwo}) imply
\begin{equation}\label{eq:bianchi1dual}
\partial_u \Theta_4+\partial_v \tilde \Theta_4=*_4 \mathcal{D}\mathcal{A}\,.
\end{equation}
From (\ref{eq:bianchi2}) and the identity above one derives
\begin{equation}\label{eq:derXi}
2\,\partial_u \partial_v \Xi= \mathcal{D}*_4 \mathcal{D}\mathcal{A} + \partial_u \mathcal{A}\wedge d\omega - \partial_v \mathcal{A}\wedge d\beta\,.
\end{equation}
Applying $\mathcal{D}$ to (\ref{eq:secondone}) and using (\ref{eq:firstone}) one obtains
\begin{equation}\label{eq:eqA0}
\begin{aligned}
\mathcal{D}*_4 \mathcal{D} *_4 \mathcal{A} =& -2 \,\partial_u\partial_v (*_4 \Xi + Z^2 \mathcal{A})\\
&=-*_4\mathcal{D}*_4 \mathcal{D}\mathcal{A} -*_4 (\partial_u \mathcal{A}\wedge d\omega - \partial_v \mathcal{A}\wedge d\beta) -2 Z^2 \partial_u\partial_v\mathcal{A}\,,
\end{aligned}
\end{equation}
where in the last step we have used (\ref{eq:derXi}). If one defines the Laplacian associated with the covariant differential $\mathcal{D}$:
\begin{equation}\label{eq:laplaciandefinition}
\nabla^2\equiv -(\mathcal{D}*_4 \mathcal{D} *_4+*_4\mathcal{D}*_4 \mathcal{D})+*_4(\partial_v \mathcal{A}\wedge d\beta-\partial_u \mathcal{A}\wedge d\omega)\,,
\end{equation}
one can prove (see Appendix \ref{sec:lapl}) that $\nabla^2$ has a simple action on forms
\begin{equation}\label{eq:laplacianidentity}
\nabla^2 = \mathcal{D}^i \mathcal{D}_i \,,
\end{equation}
where indices are contracted using the flat metric $ds^2_4$. Then eq. (\ref{eq:eqA0}) reduces to
\begin{equation}\label{eq:eqA}
\nabla^2 \mathcal{A} = 2 Z^2 \partial_u\partial_v\mathcal{A}\,,
\end{equation}
which is a set of decoupled partial differential equations for each component of the 1-form $\mathcal{A}$. These are the main dynamical equations one needs to solve to construct the linearised solution. All other gauge-invariant quantities ($Z_4$, $\Theta_4$, $\tilde \Theta_4$, $\Xi$) can be reconstructed from the 1-form $\mathcal{A}$ thanks to eqs. (\ref{eq:secondone}), (\ref{eq:bianchi1}), (\ref{eq:bianchi1dual}) and (\ref{eq:derXi}). Note that in the examples we consider in this paper the perturbation has a simple exponential dependence on $u$ and $v$, hence inverting $u$ and $v$ derivatives is a trivial task.

In summary, we need to solve eq. (\ref{eq:eqA}) with the constraint that $\mathcal{A}$ agrees with the decoupling limit result in the inner region and vanishes sufficiently fast at large distances.

\section{A non-extremal solution}
\label{sec:nonextremal}

Consider the state $[|+,+\rangle_1]^{N_1} [J^+_{-1} \tilde J^+_{-1} \,|0,0\rangle_k]^{N_2}$ in the $N_2\ll N_1$ regime: this is a ``maximally'' non-extremal perturbation of the background (\ref{eq:background},\ref{eq:background2}), where one adds energy, through the action of the currents $J^+_{-1}$ and  $\tilde J^+_{-1}$, without adding any net momentum along the $S^1$.

In the inner region, the perturbation is given by (\ref{eq:pert2charge}) with 
\begin{equation}
Y=Y^{\ell,\ell}_{-\ell+1,-\ell+1}=e^{-i(k-2)\phi} \, (k\,\cos^2\theta-1)\sin^{k-2}\theta \quad  (\mathrm{where}\,\,\ell=k/2)\,.
\end{equation}

 The inner region solution can be recast in the form (\ref{eq:pertgen}), and hence one can read off the ``near-horizon" values of the gauge-invariant quantities $Z_4$, $\mathcal{A}$, $\Theta_4$, $\tilde \Theta_4$, $\Xi$ that parametrise the perturbation. In particular we find
\begin{equation}
\mathcal{A}_\mathrm{n.h.}=e^{-i \frac{\sqrt{2}\,(u+v)}{R} - i (k-1)\phi}\, f_\mathrm{n.h.}(r,\theta)\, (dx_1+i dx_2)\,,
\end{equation}
with
\begin{equation}
f_\mathrm{n.h.}(r,\theta) =  \frac{2}{R}\,\frac{b\,a^k}{(r^2+a^2)^{\frac{k+1}{2}}}\,\sin^{k-1}\theta\,.
\end{equation}
As explained, all other gauge-invariant quantities easily follow from $\mathcal{A}$; for example
\begin{equation}
Z_{4, \mathrm{n.h.}}=R\,e^{-i \frac{\sqrt{2}\,(u+v)}{R} - i (k-2)\phi}\,\frac{b\,a^k}{(r^2+a^2)^{\frac{k}{2}}}\,\sin^{k-2}\theta\,\frac{k\,\cos^2\theta-1}{r^2+a^2\cos^2\theta}\,.
\end{equation}

A natural ansatz for the asymptotically flat extension of $\mathcal{A}$ is
\begin{eqnarray}\label{eq:AF}
\mathcal{A}=e^{-i \frac{\sqrt{2}\,(u+v)}{R} - i (k-1)\phi}\, f_\mathrm{n.h.}(r,\theta)\, f(r,\theta)\,(dx_1+i dx_2)\,,
\end{eqnarray}
where $f(r,\theta)$ is an unknown function such that $f(r,\theta)\to 1$ for $r,a\ll \sqrt{Q}$ and $\frac{f}{r^{k+1}}\to 0$ for $r\to \infty$. $f(r,\theta)$ is determined by a partial differential equation which descends from (\ref{eq:eqA}):
\begin{equation}\label{eq:eqF}
\begin{aligned}
&(r^2+a^2) \partial_r^2 f + ((1-2 k) r^2+a^2 )\frac{\partial_r f}{r}  + \partial_\theta^2 f - 2 \, \frac{ 1- 2 k \cos^2 \theta}{\sin 2\theta}  \partial_\theta f \\
& +   \frac{4}{R^2} \left[  (r^2+a^2\cos^2\theta) + 2Q \right] f = 0 \, .
\end{aligned}
\end{equation}
Note that the term in the last line is negligible for $r,a\ll \sqrt{Q}$ (which implies $Q\ll R^2$), and hence $f=1$ is a solution in the inner region, as expected. Due to the $\theta$-dependent term in the last line of eq. (\ref{eq:eqF}), that equation is not separable and thus we could not find an exact analytic solution. To provide evidence of the existence of a solution with the required boundary conditions we resort to a matched asymptotic expansion, as was done in \cite{Mathur:2003hj} and \cite{Roy:2016zzv}. This expansion applies to the regime in which one has two widely separated scales $a$ and $\sqrt{Q}$ with $a\ll \sqrt{Q}$: one can then solve the equation separately in the inner region $r\ll \sqrt{Q}$ and in the outer region $r\gg a$, and require that the two solutions match in the overlapping region $a\ll r\ll \sqrt{Q}$. We will perform the matching at leading order. We already know the inner region solution ($f=1$) so we just have to solve eq. (\ref{eq:eqF}) in the outer region.

\subsection{The solution in the outer region $r\gg a$}
When $a$ is negligible with respect to $r$ the equation for $F$ simplifies:
\begin{equation}\label{eq:eqFouter}
r^2 \partial_r^2 f + (1-2 k) r \partial_r f   +   \frac{4}{R^2} (r^2 + 2Q ) f + \partial_\theta^2 f - 2 \, \frac{ 1- 2 k \cos^2 \theta}{\sin 2\theta}  \partial_\theta f = 0 \, .
\end{equation}
This equation is separable: $f(r,\theta)=f_1(r) f_2(\theta)$; moreover, since it has to match to a constant for small $r$, we need to take a constant $f_2(\theta)$; one can also check that a constant is the only solution of the angular equation that does not have unphysical singularities for some values of $\theta$. The radial equation is a Bessel equation, whose general solution gives
\begin{equation}\label{eq:Fouter}
f(r,\theta)= r^k \left[c_1 \,J_\alpha \left(\frac{2r}{R}\right)+ c_2\, Y_\alpha \left(\frac{2r}{R}\right)\right]\quad \mathrm{with}\quad \alpha=\sqrt{k^2-\frac{8 Q}{R^2}}\,,
\end{equation}
where $c_1$ and $c_2$ are constants. Substituting this result in (\ref{eq:AF}), we see that the asymptotic behaviour of the 1-form $\mathcal{A}$ for $r\gg R$ is
\begin{equation}\label{eq:asA}
\mathcal{A}\sim \frac{1}{r^{3/2}}e^{-i \frac{\sqrt{2}\,(u+v)}{R} - i (k-1)\phi}\sin^{k-1}\theta\! \left[\tilde c_1\cos\left(\frac{2r}{R}\right)+\tilde c_2 \sin\left(\frac{2r}{R}\right)\right] \!(dx_1+i dx_2)\,,
\end{equation}
with $\tilde c_1 = c_1 \cos\left(\frac{2\alpha+1}{4}\pi\right)-c_2 \sin\left(\frac{2\alpha+1}{4}\pi\right)$, $\tilde c_2 = c_2 \cos\left(\frac{2\alpha+1}{4}\pi\right)+c_1 \sin\left(\frac{2\alpha+1}{4}\pi\right)$;
we find the same fall-off for the scalar $Z_4$: $Z_4\sim r^{-3/2}$. This is a slower fall-off than the one exhibited by the extremal solutions, and agrees with the one estimated in Section 3.3 of \cite{Mathur:2003hj} for non-extremal perturbations\footnote{Note that the same decay was found for the time-dependent non-supersymmetric solutions of \cite{Mathur:2013nja}. We thank D. Turton for pointing this out to us.}. Notice that this is a general unavoidable feature of non-extremal perturbations, since all non-trivial solutions of eq. (\ref{eq:eqA}) have this fall-off; the only way to a obtain a faster asymptotic decay is to impose $\mathcal{A}=0$, which implies that the perturbation is $u$ and/or $v$-independent, i.e. it is extremal. We will show that, despite the slowness of the fall-off, the global charges of the solution are not altered by the perturabation, and thus the solution can be consistently identified with a microstate of a D1D5P black hole. 

\subsection{The matching region $a\ll r \ll \sqrt{Q}$}
Consistency with the near-horizon solution requires that in the limit $r\ll\sqrt{Q}$ (and $a\ll \sqrt{Q}$) the function (\ref{eq:Fouter}) tends to 1, for some choice of the constants $c_i$. This actually guaranteed a priori, since the asymptotic analysis has not imposed any constraint on the integration constants $c_i$, and we know that the equation for $f$ has the solution $f=1$ in the inner region. Indeed one finds in the small $r$ limit
\begin{equation}
f(r,\theta)\approx r^k \left[\frac{c_1}{k!}\left(\frac{r}{R}\right)^k - \frac{c_2 \,(k-1)!}{\pi}\,\left(\frac{r}{R}\right)^{-k} \right]\,,
\end{equation}
where we have approximated $\alpha \approx k$ since $Q\ll R^2$ for $a\ll \sqrt{Q}$. Hence the solution matches at leading order if $c_2=-\pi\,R^{-k}/(k-1)!$.

\subsection{Asymptotic charges}
We have shown the existence of a solution which interpolates between the near-horizon and the asymptotic regions; the fields of the perturbation fall off at large distances like $r^{-3/2}$. This is a very slow decay: in 5 non-compact dimensions a field strength carrying a global charge vanishes like $r^{-3}$, and we expect our perturbation to decay faster, so as to leave the global charges of the solution invariant. We show here that the unusually slow decay is in fact not a problem, as the non-trivial angular dependence of the perturbation guarantees that it does not contribute to the global charges.

Since the perturbation excites the B-field, it could carry a global F1 and NS5 charge, proportional to 
\begin{equation}
Q_{F1} \sim \int_{S^3} *_6 H \,,\quad Q_{NS5} \sim \int_{S^3} H  \,,
\end{equation}
where the integral is over a 3-sphere with infinite radius in the four non-compact spatial directions. It follows from eq. (\ref{eq:first}), and the fact that the $w F$ term is negligible at large $r$, that
\begin{equation}
\int_{S^3} H = \int_{S^3} *_6 H = \int_{S^3} \Xi\,.
\end{equation}
The large $r$ limit of the 3-form $\Xi$ can be computed from the asymptotic expression for $\mathcal{A}$ in (\ref{eq:asA}) via eq. (\ref{eq:derXi}), where one can discard the last two terms at large distances:
\begin{equation}
2\,\partial_u \partial_v\, \Xi\approx \mathcal{D}*_4 \mathcal{D} \mathcal{A}\,. 
\end{equation}
One finds
\begin{equation}
\begin{aligned}
& \int_{S^3} \Xi \sim \lim_{r\to\infty} r^{1/2}\left[\tilde c_1\sin\left(\frac{2r}{R}\right)-\tilde c_2 \cos\left(\frac{2r}{R}\right)\right]\,\times\\
&\times\,\int \!d\theta\, d\phi\, d\psi\,e^{-i \frac{\sqrt{2}\,(u+v)}{R} - i (k-2)\phi}\,\sin^{k-1}\theta \cos\theta\,[(k+2)\cos2\theta +3 (k-2)]\,.
 \end{aligned}
\end{equation}
Based only on the $r$-dependence of $\Xi$ one would conclude that the charge carried by the perturbation is not only non-vanishing, but divergent. However the integral over the angular variables vanishes for any $k>0$: when $k\not=2$ the oscillating factor $e^{-i(k-2)\phi}$ kills the $\phi$-integral, while for $k=2$ it is the $\theta$ integral that vanishes: $\int_0^{\pi/2} d\theta \sin (4\theta)=0$. Note that the state with $k=2$ is special because it does not depend on either $\phi$ or $\psi$: this is a consequence of the fact that the strand $J^+_{-1} \tilde J^+_{-1} \,|0,0\rangle_2$ carries the same angular momenta as $(|\!++\rangle_1)^2$.

\section{A near-extremal solution}
\label{sec:nearextremal}
We want to consider here a non-supersymmetric state where the departure from extremality could be made arbitrarily small. We could start from the supersymmetric D1D5P state $[|+,+\rangle_1]^{N_1} [(J^+_{-1})^k |0,0\rangle_k]^{N_2}$: note that $k$ is the maximum number of times the charge $J^+_{-1}$ can act on the ground state $|0,0\rangle_k$, since $(J^+_{-1})^{k+1} |0,0\rangle_k=0$. We can break supersymmetry by acting once with the right-moving current $\tilde J^+_{-1}$: $[|+,+\rangle_1]^{N_1} [(J^+_{-1})^k \tilde J^+_{-1} |0,0\rangle_k]^{N_2}$. In the limit of large $k$ one would expect this to be a small perturbation of the supersymmetric state: we will thus set-up a large $k$ expansion and keep the first non-trivial order in $1/k$. As usual we will also assume $N_2\ll N_1$, so we can linearise the supergravity equations around the background (\ref{eq:background},\ref{eq:background2}).

The solution in the inner region is given by (\ref{eq:pert2charge}) with 
\begin{equation}
Y = Y^{\ell,\ell}_{\ell,-\ell+1} = e^{i(k-1)\psi+i\phi}\cos^{k-1}\!\theta\, \sin\theta\quad (\ell=k/2)\,,
\end{equation}
where we discard the spherical harmonic normalisation factor. We can extract from this solution the near-horizon values of the functions that appear in the ansatz (\ref{eq:pertgen}); for example:
\begin{subequations}
\begin{align}
Z_{4,\mathrm{n.h.}}&=R\,b\,e^{-i\frac{\sqrt{2}}{R}(u+k v)+i(k-1)\psi+i\phi}\,\frac{\Delta_{k,k-1}}{\Sigma}\,,\\ 
\mathcal{A}_\mathrm{n.h.}&=-2\frac{b}{R}\,e^{-i\frac{\sqrt{2}}{R}(u+k v)+i(k-1)\psi}\,\frac{\Delta_{k,k-1}}{\sqrt{r^2+a^2}\sin\theta}\,(dx_1+i dx_2)\,,\label{eq:Anh}
\end{align}
\end{subequations}  
where we define
\begin{equation}
\Delta_{k,m}\equiv \left(\frac{a}{\sqrt{r^2+a^2}}\right)^{k}\sin^{k-m}\theta \cos^m\theta\,.
\end{equation}
This is an exact solution of the equations of motion in the inner region for any $k$. 

We expect that the problem of extending the solution outside of the inner region simplifies in the regime of large $k$, in which the state becomes approximately extremal. For this reason we look for a solution of the equations of motion (\ref{eq:firstbis}), (\ref{eq:secondone}) in the form of an expansion in $1/k$, and only keep the first non-trivial order:
\begin{equation}
\begin{aligned}
Z_4 &= Z_{4,0}+k^{-1} Z_{4,1} + \mathcal{O}(k^{-2})\,,\quad \mathcal{A} = \mathcal{A}_{0}+k^{-1} \mathcal{A}_{1} + \mathcal{O}(k^{-2})\,.\\
\end{aligned}
\end{equation}
In defining the large $k$ expansion, we keep the exact $k$-dependence of exponents, so we do not expand the oscillating factor $\exp [-i\frac{\sqrt{2}}{R}(u+k v)+i(k-1)\psi+i\phi]$ or $\Delta_{k,k-1}$, and only expand the $k$-dependent coefficients that multiply the various functions. According to this definition, $Z_{4}, \mathcal{A}, \tilde\Theta_4$ start at order $k^0$, while the leading term of $\Theta_4$ is of order $k^1$. Moreover, when $v$, $r$, $\theta$ and $\psi$ derivatives act on our solution, they increase the $k$-order by one (as a consequence of the $k$-dependence of $\exp [-i\frac{\sqrt{2}}{R}(u+k v)+i(k-1)\psi+i\phi]$ and $\Delta_{k,k-1}$), while $u$ and $\phi$ derivatives do not change the order in $k$: schematically $\mathcal{D}, \partial_v \sim k^1$, $\partial_u\sim k^0$.

One can now see how the equations of motion simplify at large $k$. As explained in Section~\ref{sec:equations} it is convenient to derive $\mathcal{A}$ using eq. (\ref{eq:eqA}); the remaining gauge invariant quantities follow from $\mathcal{A}$ without the need to integrate any further differential equation. The leading contribution to the l.h.s. of (\ref{eq:eqA}) is of order $k^2$ (as $\nabla^2\sim k^2$) , while the r.h.s. starts at order $k$. Hence at leading order one should require
\begin{equation}
\nabla^2 \mathcal{A}_0 = \mathcal{O}(k)\,.
\end{equation}
Since $Z$ has disappeared from the equation above, the solution for $\mathcal{A}$ at leading order in $k$ coincides with the near-horizon solution even outside the inner region:
\begin{equation}
\mathcal{A}_{0}= \mathcal{A}_{\mathrm{n.h.}}\,.
\end{equation}
At the next order in $1/k$, the l.h.s. of (\ref{eq:eqA}) has two contributions: the order leading contribution to $k^{-1} \nabla^2 \mathcal{A}_1$, and the order $k$ contribution to $\nabla^2 \mathcal{A}_\mathrm{n.h.}$, which is given by
\begin{equation}
\nabla^2 \mathcal{A}_\mathrm{n.h.} = 2\, \frac{Q^2}{\Sigma^2} \,\partial_u \partial_v \mathcal{A}_\mathrm{n.h.}\,,
\end{equation}
as a consequence of the near-horizon equations of motion. On the r.h.s. one can approximate $\mathcal{A}$ with $\mathcal{A}_\mathrm{n.h.}$ up to corrections of $\mathcal{O}(k^0)$. Hence the first non-extremal correction to our solution is determined by
\begin{equation}\label{eq:eqA1}
k^{-1} \nabla^2 \mathcal{A}_1 = 2 \left(1+2\frac{Q}{\Sigma}\right) \partial_u \partial_v \mathcal{A}_\mathrm{n.h.}+\mathcal{O}(k^0)\,.
\end{equation}
Given the form of $\mathcal{A}_\mathrm{n.h.}$ (\ref{eq:Anh}), one can search a solution for $\mathcal{A}_1$ of the form
\begin{equation}
\mathcal{A}_1 = 2 \frac{b}{R}\,e^{-i\frac{\sqrt{2}}{R}(u+k v)+i(k-1)\psi}\,G(r,\theta)\,(dx_1+i dx_2)\,.
\end{equation}
Then (\ref{eq:eqA1}) implies
\begin{equation}\label{eq:poisson}
\widehat{\mathcal{L}}^{(k,k)} G=\frac{4 k^2}{R^2} \left(1+2\frac{Q}{\Sigma}\right)\,\frac{\Delta_{k,k-1}}{\sqrt{r^2+a^2}\sin\theta}+\mathcal{O}(k)\,,
\end{equation}
where $\widehat{\mathcal{L}}^{(k,k)}$ is the covariant Laplacian that was defined in \cite{Bena:2015bea}:
 \begin{equation}
 \begin{aligned}
 \widehat{\mathcal{L}}^{(k,k)} G &\equiv \frac{1}{r \Sigma} \partial_r(r(r^2+a^2) \partial_r G)+\frac{1}{\Sigma\sin\theta\cos\theta}\partial_\theta(\sin\theta\cos\theta\, \partial_\theta G)-k^2 \frac{r^2+a^2 \sin^2\theta}{(r^2+a^2) \Sigma\cos^2\theta}G\\
&\approx \frac{r^2+a^2}{\Sigma} \partial^2_r G+\frac{1}{\Sigma}\partial^2_\theta G-k^2 \frac{r^2+a^2 \sin^2\theta}{(r^2+a^2) \Sigma\cos^2\theta}G \,,
\end{aligned}
 \end{equation}
where in the second line we have kept only the terms of order $k^2$, according to our working assumption that $r$ and $\theta$ derivatives of $G$ give terms of order $k$. 

We have thus reduced our problem to the solution of a Poisson equation for the deformed Laplacian $\widehat{\mathcal{L}}^{(k,k)} G$; equations of this type appeared in the construction of extremal superstrata \cite{Bena:2015bea}, but the source term in (\ref{eq:poisson}) is different from the one of \cite{Bena:2015bea}. Though we do not exclude that a variation of the techniques of \cite{Bena:2015bea} could be used to find an exact solution of (\ref{eq:poisson}), we have not been able to find one. Hence we resort to a matching technique to show that (\ref{eq:poisson}) admits a solution that is well behaved at large distances and is negligible with respect to $\mathcal{A}_\mathrm{n.h.}$ in the inner region.

\subsection{A matching near-extremal solution}
We assume as usual that $a\ll \sqrt{Q}$ and look for a solution in the outer region $r\gg a$, where the l.h.s. of (\ref{eq:poisson}) approximates to
\begin{equation}\label{eq:lhsouter}
 \widehat{\mathcal{L}}^{(k,k)} G \approx \partial^2_r G+\frac{1}{r^2}\partial^2_\theta G- \frac{k^2}{r^2 \cos^2\theta}G +\frac{a^2}{r^2}\left[\sin^2\theta\,\partial^2_r G-\frac{\cos^2\theta}{r^2}\,\partial^2_\theta G+\frac{2 k^2}{r^2}G\right]
\end{equation}
and the r.h.s. to
\begin{equation}\label{eq:rhsouter}
\mathrm{r.h.s.}\approx \frac{4 k^2}{R^2}\left(1+\frac{2 Q}{r^2}\right)\,\frac{a^k}{r^{k+1}}\,\cos^{k-1}\theta\,.
\end{equation}
One can look for a factorised solution of the form
\begin{equation}\label{eq:poissonguess}
G(r,\theta) \approx g(\theta)\left(1+\frac{2 Q}{r^2}\right)\, \frac{a^k}{r^{k+n}}\,\cos^{k-1}\theta\,,
\end{equation}
where $n$ is a number that we assume to be much smaller than $k$ ($n\ll k$) and that will be determined shortly. At leading order in $1/k$, one can approximates $\partial_r^2 G \approx k^2/r^2 \,G$ and $\partial_\theta^2 G \approx k^2 \tan^2\theta\,G$, so that when one substitutes (\ref{eq:poissonguess}) into (\ref{eq:lhsouter}) one immediately sees that the leading term in $a/r$ vanishes for any choice of $n$ up to terms of $\mathcal{O}(k)$:
\begin{equation}
 \partial^2_r G+\frac{1}{r^2}\partial^2_\theta G- \frac{k^2}{r^2 \cos^2\theta}G = \frac{k^2}{r^2} \left[1+\tan^2\theta-\frac{1}{\cos^2\theta}\right]G+ \mathcal{O}(k)=\mathcal{O}(k)\,.
\end{equation}
Hence only the term proportional to $a^2/r^2$ survives in (\ref{eq:lhsouter}) and to match the source (\ref{eq:rhsouter}) one needs $n=-3$. The equation for $G$ then becomes
\begin{equation}
\frac{k^2\, a^2}{r^4} [\sin^2\theta - \sin^2\theta +2 ] G = \frac{4 k^2}{R^2}\left(1+\frac{2 Q}{r^2}\right)\,\frac{a^k}{r^{k+1}}\,\cos^{k-1}\theta + \mathcal{O}(k)\,,
\end{equation}
which is satisfied by taking
\begin{equation}
g(\theta) = \frac{2}{Q^2}\,.
\end{equation}
Then the solution for $\mathcal{A}_1$ in the outer region is
\begin{equation}
\mathcal{A}_1 \approx  4 \frac{b}{Q^2 R}\,e^{-i\frac{\sqrt{2}}{R}(u+k v)+i(k-1)\psi}\,\left(1+\frac{2 Q}{r^2}\right)\, \frac{a^k}{r^{k-3}}\,\cos^{k-1}\theta\,(dx_1+i dx_2)\,.
\end{equation}
Consistency requires that in the matching region ($a\ll r \ll \sqrt{Q}$) $\mathcal{A}_1$ be suppressed with respect to $\mathcal{A}_\mathrm{n.h.}$; this is evidently so, as
\begin{equation}
\frac{|\mathcal{A}_1|}{|\mathcal{A}_\mathrm{n.h.}|}\sim \frac{r^4}{Q^2}\ll 1 \quad \mathrm{for}\quad a\ll r \ll \sqrt{Q}\,.
\end{equation}
The remaining fields in the outer region can be reconstructed from $\mathcal{A}_1$; for example:
\begin{equation}
Z_{4,1}\approx -2 \frac{R\,b}{Q^2}\,e^{-i\frac{\sqrt{2}}{R}(u+k v)+i(k-1)\psi+i\phi}\,\left(1+\frac{2 Q}{r^2}\right)\, \frac{a^k}{r^{k-2}}\,\cos^{k-1}\theta \sin\theta\,.
\end{equation} 
One can also see that in the near-extremal regime $k\gg 1$ the fields of the perturbation fall-off very fast at large distances (for example $Z_4\sim 1/r^{k-2}$). This is to be contrasted with the much slower ($r^{-3/2}$) fall-off seen for the non-extremal solution. This indicates that the $k\to \infty$ and $r\to \infty$ limits do not commute: our near-extremal expansion is valid up to a distance $r$ that grows with $k$, while for larger distances one should recover the large $r$ behaviour of non-extremal solutions. 

\section{Summary and outlook}
We have constructed linearised solutions of the type IIB equations of motion that are dual to non-extremal states of the D1D5P system obtained by acting with left and right-moving chiral algebra generators on a 1/4 BPS state. We have used an ansatz inspired by the supersymmetric geometries. However we have shown that already at the linear level the solution of the equations is significantly more difficult for non-extremal configurations. In our ansatz the complication arises from the fact that the warp factor $Z$ does not decouple from the equations for the perturbation: hence the problem of extending the solution outside of the inner region requires solving a non-trivial differential problem. The problem is sourced by the 1-form $\mathcal{A}$, which couples to $Z$ through the last term of eq. (\ref{eq:firstone}); it is evident from eq. (\ref{eq:secondone}) that $\mathcal{A}$ does not vanish precisely when the perturbation is non-extremal and depends on both light-cone coordinates $u$ and $v$. From the technical point of view, our main result is the reduction of the differential problem to a single equation (\ref{eq:eqA}) of the Poisson type for $\mathcal{A}$. The full perturbation can be reconstructed from $\mathcal{A}$ without having to solve any further differential equation. Though we have not been able to solve the $\mathcal{A}$ equation exactly, we have shown that it admits a solution that interpolates between the inner region result and an asymptotically decaying solution. Despite the unusually slow fall off of the solution at large distances, the global charges of the solution are well defined and equal to the ones of the D1D5P black hole, supporting the identification of our solutions with black hole microstates. We have also developed an approximation scheme that allows to consistently expand near-extremal solutions around a supersymmetric background.

The most obvious development of our work would be the extension of our perturbative solutions to fully non-linear solutions of the supergravity equations. This would be interesting not only as a technical achievement but it would also have a significant conceptual spin-off, as it would imply the existence of a non-supersymmetric analogue of the supergraviton gas. We recall that the geometries dual to the supergraviton gas states were constructed in \cite{Bena:2015bea} by first taking linear combinations of the linear solutions corresponding to superdescendants and then building up the non-linear complication of the solution. The existence of the non-linear solution heavily relied on the linear structure of the BPS equations and also on the fact that the spatial 4D base $ds^2_4$ did not receive non-linear corrections. It is thus highly non-obvious that a similar method could be used in the non-supersymmetric case. A preliminary analysis of the non-linear non-extremal solution generated in the decoupling limit via the action of the chiral algebra indeed indicates that $ds^2_4$ receives corrections at non-linear orders. Probably a more manageable problem is the non-linear completion of the near-extremal solution of Section \ref{sec:nonextremal}: we believe that a non-linear extension of the large $k$ expansion should be feasible and we hope to return on this in future work. 

\vspace{7mm}
 \noindent {\large \textbf{Acknowledgements} }

 \vspace{5mm} 

We would like to thank S. Mathur, R. Russo and D. Turton for useful discussions and correspondence. We thank the Galileo Galilei Institute for Theoretical Physics (GGI) for the hospitality during the program ``New Developments in AdS3/CFT2 Holography''. This research is partially supported by the Padova University Project CPDA119349 and by INFN.

\appendix

\section{Supergravity equations}
\label{sec:appeqs}
We show here in detail how eqs. (\ref{eq:firstbis}, \ref{eq:secondone}) follow from the linearised equations of motion (\ref{eq:motionlinear}). We first express the field strengths $H$, $*H$ and $F$ in terms of the quantities that appear in the general ansatz  \eqref{eq:background},\eqref{eq:pertgen}. To simplify the notation we define 
\begin{equation}
d\hat u \equiv du+\omega, \quad d\hat v \equiv dv+\beta\,;
\end{equation}
note that these are not exact differentials, but $d(d\hat u) = d\omega, \, d(d\hat v) = d\beta$. We find
\begin{subequations}
\begin{align}
H = \, &  -\left[ \mathcal{D} \left( \frac{Z_4}{Z^2} \right) + \mathcal{A}  \right] \wedge d\hat u \wedge d\hat v +  \left[ \Theta_4 -  \frac{Z_4}{Z^2} d \omega \right] \wedge d\hat v + \left[\tilde \Theta_4 +  \frac{Z_4}{Z^2} d \beta \right] \wedge d\hat u + \Xi \,,\\
*_6 H  = \, &  \frac{1}{Z^2} *_4 \Xi  \wedge d\hat u \wedge d\hat v  + \left[*_4 \Theta_4 +  \frac{Z_4}{Z^2} d \omega \right] \wedge d\hat v - \left[*_4 \tilde \Theta_4 +  \frac{Z_4}{Z^2} d \beta \right] \wedge d\hat u  \nonumber\\
&+ Z^2 *_4 \left[  \mathcal{D} \left( \frac{Z_4}{Z^2} \right) + \mathcal{A} \right] \,,\\
F = \, &  \frac{d Z}{Z^2} \wedge d\hat u \wedge d\hat v - \frac{1}{Z}  \left[ d \omega \wedge d\hat v - d \beta \wedge d\hat u \right] + *_4 dZ\,.
\end{align}
\end{subequations}
In writing $F$ we have used that $d\gamma=*_4 dZ$.

Eq. (\ref{eq:firstzero}) can be integrated to (\ref{eq:first}), whose two terms are given by
\begin{equation}
\begin{split}
*_6 H - H =\, & \left[ \frac{ \mathcal{D} Z_4 }{ Z^2 } + \mathcal{A} + \frac{1}{ Z^2 } *_4\Xi   - 2\frac{Z_4}{Z^3} d Z \right]  \wedge d\hat u \wedge d\hat v \\ 
& -\left[ \left( \Theta_4 - *_4 \Theta_4 \right) - 2 \frac{Z_4}{Z^2} d \omega \right] \wedge d\hat v -\left[ \left( \tilde \Theta_4 + *_4 \tilde \Theta_4 \right) + 2 \frac{Z_4}{Z^2} d \beta \right] \wedge d\hat u \\
& + \left[ -\Xi     + Z^2 *_4 \mathcal{A} +*_4 \mathcal{D} Z_4  - 2 \frac{Z_4}{Z^3} d Z \right]
\end{split}
\end{equation}
and
\begin{equation}
2 w F = 2 \frac{Z_4}{Z^3} d Z \wedge d\hat u \wedge d\hat v - 2 \frac{Z_4}{Z^2} \left[ d \omega \wedge d\hat v - d \beta \wedge d\hat u \right] + 2 \frac{Z_4}{Z} *_4 d Z\,.
\end{equation}
Hence eq. (\ref{eq:first}) reads
\begin{equation}
\begin{split}
0=\, &  \frac{1}{ Z^2 } \left[ \mathcal{D} Z_4+Z^2  \mathcal{A} + *_4 \Xi  \right]  \wedge d\hat u \wedge d\hat v  + \left[ - \Xi   + Z^2 *_4 \mathcal{A} +*_4 \mathcal{D} Z_4  \right]\\
& -  \left( \Theta_4 - *_4 \Theta_4 \right)  \wedge d\hat v  -  \left( \tilde \Theta_4 + *_4 \tilde \Theta_4 \right)  \wedge d\hat u\,,
\end{split}
\end{equation}
which is equivalent to the relations (\ref{eq:firstbis}). 

As for eq. (\ref{eq:second}), one has
\begin{equation}\label{eq:secondpart1}
\begin{split}
d* d w =\, &  d* \left[ \frac{\mathcal{D} Z_4}{Z} - \frac{Z_4}{Z^2} d Z  + \frac{1}{Z} \partial_u Z_4 d\hat u + \frac{1}{Z} \partial_v Z_4 d\hat v \right] \\
=   \, & d\left[ \left( \frac{1}{Z} *_4 \mathcal{D} Z_4 - \frac{Z_4}{Z^2} *_4 d Z\right) \wedge d\hat u \wedge d\hat v   - Z \left( \partial_u Z_4 d\hat u - \partial_v Z_4 d\hat v\right) \wedge \mathrm{vol}_4     \right]  \\
=\, & \left[ \frac{1}{Z} \mathcal{D}*_4 \mathcal{D} Z_4  -2 \frac{\mathcal{D} Z_4}{Z^2} \wedge *_4 d Z + \frac{2 Z_4}{Z^3} d Z \wedge *_4 d Z \right. \\
& \left.  \quad - \frac{Z_4}{Z^2}d *_4 d Z  + 2 Z \partial_u \partial_v Z_4 \mathrm{vol}_4 \right] \wedge d\hat u \wedge d\hat v \\
=\, & \left[- \frac{1}{Z} \Theta_4 \wedge d \beta -\frac{1}{Z} \tilde \Theta_4 \wedge d \omega - \frac{1}{Z} \mathcal{D} \left( Z^2 *_4 \mathcal{A} \right) 
-2 \frac{\mathcal{D} Z_4}{Z^2} \wedge *_4 d Z  \right. \\
& \left.  \quad + \frac{2 Z_4}{Z^3} d Z \wedge *_4 d Z+ 2 Z \partial_u \partial_v Z_4 \mathrm{vol}_4 \right] \wedge d\hat u \wedge d\hat v \,,
\end{split}
\end{equation}
where $\mathrm{vol}_4$ is the volume of $ds^2_4$, we have used the identity 
\begin{equation}\label{eq:identitystar}
\omega_1 \wedge *_4\, \omega_2 = \omega_2 \wedge *_4 \,\omega_1 = -*_4 \omega_1 \wedge  \omega_2  
\end{equation}
valid for any 1-forms $\omega_1$, $\omega_2$, the fact that $Z$ is a harmonic function $d *_4 d Z=0$ and the equation for $Z_4$:
\begin{equation}
 \mathcal{D} *_4 \mathcal{D} Z_4  = - \Theta_4 \wedge d \beta - \tilde \Theta_4 \wedge d \omega - \mathcal{D} \left( Z^2 *_4 \mathcal{A} \right) \,,
\end{equation}
which follows by applying $\mathcal{D}$ to (\ref{eq:firstone}) and using the Bianchi identity (\ref{eq:bianchi3}). Moreover
\begin{equation}\label{eq:secondpart2}
\begin{split}
F\wedge H = & \, \left[ \frac{d Z}{Z^2} \wedge \Xi + *_4 d Z \wedge \left(2 \frac{Z_4}{Z^3} d Z -\frac{\mathcal{D} Z_4}{Z^2} -\mathcal{A} \right) \right.  \\
& \left. + \frac{1}{Z} \left(  d \omega \wedge  \tilde \Theta_4   + d \beta \wedge  \Theta_4  \right) \right] \wedge d\hat u \wedge d \hat v\\
=&\, \left[ \frac{d Z}{Z^2} \wedge \left(\Xi  -2 \frac{Z_4}{Z} *_4 d Z +*_4\mathcal{D} Z_4 +Z^2 *_4\mathcal{A} \right) \right.  \\
& \left. + \frac{1}{Z} \left(  d \omega \wedge  \tilde \Theta_4   + d \beta \wedge  \Theta_4  \right) \right] \wedge d\hat u \wedge d \hat v\\
=&\, \left[ \frac{d Z}{Z^2} \wedge \left(-2 \frac{Z_4}{Z} *_4 d Z +2*_4\mathcal{D} Z_4 +2Z^2 *_4\mathcal{A} \right) \right.  \\
& \left. + \frac{1}{Z} \left(  d \omega \wedge  \tilde \Theta_4   + d \beta \wedge  \Theta_4  \right) \right] \wedge d\hat u \wedge d \hat v\,,
\end{split}
\end{equation}
where we have again used (\ref{eq:identitystar}) and (\ref{eq:firstone}).  Eq. (\ref{eq:second}) then follows by summing (\ref{eq:secondpart1}) and (\ref{eq:secondpart2}), which yields 
\begin{equation}
\begin{split}
0=&\,- \frac{1}{Z} \mathcal{D} \left( Z^2 *_4 \mathcal{A} \right) +2 Z \partial_u \partial_v Z_4 \mathrm{vol}_4 + 2 \,dZ\wedge *_4 \mathcal{A}\\
=&\,Z\,(-\mathcal{D}*_4 \mathcal{A}+2\, \partial_u \partial_v Z_4 \mathrm{vol}_4)\,,
\end{split}
\end{equation}
whose Hodge dual is equivalent to (\ref{eq:secondone}).

\section{The $\mathcal{D}$-Laplacian}
\label{sec:lapl}
In this appendix we prove that the ``covariant Laplacian" $\nabla^2$ defined in (\ref{eq:laplaciandefinition}) has the simple action (\ref{eq:laplacianidentity}) on forms expressed in cartesian coordinates. Let us first look at the operator 
\begin{equation}
\Delta \mathcal{A} \equiv -(\mathcal{D} *_4 \mathcal{D} *_4 \mathcal{A} + *_4 \mathcal{D} *_4 \mathcal{D} \mathcal{A})\,.
\end{equation}
Writing down the single terms in components one finds
\begin{equation}
\begin{split}
\mathcal{D} *_4 \mathcal{D} *_4 \mathcal{A} &= - \mathcal{D} *_4 \mathcal{D} \left( \mathcal{A}^i\, \varepsilon_{ijkl} \,dx^j \wedge dx^k \wedge dx^l \right) \\
&= -dx^i \mathcal{D}_i \mathcal{D}_j \mathcal{A}^j , \\
*_4 \mathcal{D} *_4 \mathcal{D} \mathcal{A} &=  *_4 \mathcal{D} *_4 \left( \mathcal{D}_i \mathcal{A}_j\, dx^i \wedge dx^j  \right) \\
&= - dx^i \left(  \mathcal{D}^j \mathcal{D}_j \mathcal{A}_i - \mathcal{D}_j \mathcal{D}_i \mathcal{A}^j \right)\,,
\end{split}
\end{equation}
where indices are contracted with the flat $\mathbb{R}^4$ metric $ds^2_4$. Then
\begin{equation}
\Delta \mathcal{A} =  dx^i \, \mathcal{D}^j \mathcal{D}_j \mathcal{A}_i + dx^i \left( \mathcal{D}_i \mathcal{D}_j - \mathcal{D}_j \mathcal{D}_i \right) \mathcal{A}^j.
\end{equation}
The last term can be simplified to
\begin{equation}
\begin{split}
 dx^i \left( \mathcal{D}_i \mathcal{D}_j - \mathcal{D}_j \mathcal{D}_i \right) \mathcal{A}^j &= - dx^i (\partial_i \beta_j - \partial_j \beta_i)\partial_v \mathcal{A}^j -  dx^i (\partial_i \omega_j - \partial_j \omega_i )\partial_u \mathcal{A}^j\\
 &= - *_4 (d\beta \wedge \partial_v \mathcal{A})  + *_4 (d\omega \wedge \partial_u \mathcal{A}) \,,
\end{split}
\end{equation}
where we have used that
\begin{equation}
\mathcal{D}^2 = -d\beta\wedge \partial_v -d\omega\wedge \partial_u\,,
\end{equation}
and the (anti-)self-duality of $d\beta$ ($d\omega$): $d\beta=*_4d\beta$, $d\omega=-*_4d\omega$. So finally we find
\begin{equation}
\Delta \mathcal{A} = dx^j\,\mathcal{D}^i \mathcal{D}_i \mathcal{A}_j -   *_4 \left( \partial_v \mathcal{A} \wedge d \beta - \partial_u \mathcal{A} \wedge d \omega \right) \,.
\end{equation}
Then the definition (\ref{eq:laplaciandefinition}) gives
\begin{equation}
\nabla^2 = \Delta +   *_4 \left( \partial_v \mathcal{A} \wedge d \beta - \partial_u \mathcal{A} \wedge d \omega \right) =  \mathcal{D}^i \mathcal{D}_i\,,
\end{equation}
which proves (\ref{eq:laplacianidentity}).

\providecommand{\href}[2]{#2}\begingroup\raggedright\endgroup



\begin{thebibliography}{10}

\bibitem{Strominger:1996sh}
A.~Strominger and C.~Vafa, ``{Microscopic origin of the Bekenstein-Hawking
  entropy},'' {\em Phys.Lett.} {\bf B379} (1996) 99--104,
  \href{http://arXiv.org/abs/hep-th/9601029}{{\tt hep-th/9601029}}.

\bibitem{Dijkgraaf:1996xw}
R.~Dijkgraaf, G.~W. Moore, E.~P. Verlinde, and H.~L. Verlinde, ``{Elliptic
  genera of symmetric products and second quantized strings},'' {\em Commun.
  Math. Phys.} {\bf 185} (1997) 197--209,
\href{http://arXiv.org/abs/hep-th/9608096}{{\tt hep-th/9608096}}.

\bibitem{Maldacena:1999bp}
J.~M. Maldacena, G.~W. Moore, and A.~Strominger, ``{Counting BPS black holes in
  toroidal Type II string theory},''
\href{http://arXiv.org/abs/hep-th/9903163}{{\tt hep-th/9903163}}.

\bibitem{Kanitscheider:2006zf}
I.~Kanitscheider, K.~Skenderis, and M.~Taylor, ``{Holographic anatomy of
  fuzzballs},'' {\em JHEP} {\bf 0704} (2007) 023,
\href{http://arXiv.org/abs/hep-th/0611171}{{\tt hep-th/0611171}}.

\bibitem{Kanitscheider:2007wq}
I.~Kanitscheider, K.~Skenderis, and M.~Taylor, ``{Fuzzballs with internal
  excitations},'' {\em JHEP} {\bf 06} (2007) 056,
\href{http://arXiv.org/abs/0704.0690}{{\tt 0704.0690}}.

\bibitem{Giusto:2015dfa}
S.~Giusto, E.~Moscato, and R.~Russo, ``{AdS$_{3}$ holography for 1/4 and 1/8
  BPS geometries},'' {\em JHEP} {\bf 11} (2015) 004,
\href{http://arXiv.org/abs/1507.00945}{{\tt 1507.00945}}.

\bibitem{Lunin:2001dt}
O.~Lunin and S.~D. Mathur, ``{The slowly rotating near extremal D1-D5 system as
  a 'hot tube'},'' {\em Nucl. Phys.} {\bf B615} (2001) 285--312,
\href{http://arXiv.org/abs/hep-th/0107113}{{\tt hep-th/0107113}}.

\bibitem{Avery:2009tu}
S.~G. Avery, B.~D. Chowdhury, and S.~D. Mathur, ``{Emission from the D1D5
  CFT},'' {\em JHEP} {\bf 10} (2009) 065,
\href{http://arXiv.org/abs/0906.2015}{{\tt 0906.2015}}.

\bibitem{Balasubramanian:2011ur}
V.~Balasubramanian, A.~Bernamonti, J.~de~Boer, N.~Copland, B.~Craps,
  E.~Keski-Vakkuri, B.~Muller, A.~Schafer, M.~Shigemori, and W.~Staessens,
  ``{Holographic Thermalization},'' {\em Phys. Rev.} {\bf D84} (2011) 026010,
\href{http://arXiv.org/abs/1103.2683}{{\tt 1103.2683}}.

\bibitem{Maldacena:2015waa}
J.~Maldacena, S.~H. Shenker, and D.~Stanford, ``{A bound on chaos},'' {\em
  JHEP} {\bf 08} (2016) 106,
\href{http://arXiv.org/abs/1503.01409}{{\tt 1503.01409}}.

\bibitem{Avery:2010er}
S.~G. Avery, B.~D. Chowdhury, and S.~D. Mathur, ``{Deforming the D1D5 CFT away
  from the orbifold point},'' {\em JHEP} {\bf 06} (2010) 031,
\href{http://arXiv.org/abs/1002.3132}{{\tt 1002.3132}}.

\bibitem{Carson:2014ena}
Z.~Carson, S.~Hampton, S.~D. Mathur, and D.~Turton, ``{Effect of the
  deformation operator in the D1D5 CFT},'' {\em JHEP} {\bf 1501} (2015) 071,
\href{http://arXiv.org/abs/1410.4543}{{\tt 1410.4543}}.

\bibitem{Burrington:2014yia}
B.~A. Burrington, S.~D. Mathur, A.~W. Peet, and I.~G. Zadeh, ``{Analyzing the
  squeezed state generated by a twist deformation},'' {\em Phys.Rev.} {\bf D91}
  (2015), no.~12, 124072,
\href{http://arXiv.org/abs/1410.5790}{{\tt 1410.5790}}.

\bibitem{Mathur:2005zp}
S.~D. Mathur, ``{The fuzzball proposal for black holes: An elementary
  review},'' {\em Fortsch. Phys.} {\bf 53} (2005) 793--827,
\href{http://arXiv.org/abs/hep-th/0502050}{{\tt hep-th/0502050}}.

\bibitem{Skenderis:2008qn}
K.~Skenderis and M.~Taylor, ``{The fuzzball proposal for black holes},'' {\em
  Phys. Rept.} {\bf 467} (2008) 117--171,
\href{http://arXiv.org/abs/0804.0552}{{\tt 0804.0552}}.

\bibitem{Balasubramanian:2008da}
V.~Balasubramanian, J.~de~Boer, S.~El-Showk, and I.~Messamah, ``{Black Holes as
  Effective Geometries},'' {\em Class.Quant.Grav.} {\bf 25} (2008) 214004,
  \href{http://arXiv.org/abs/0811.0263}{{\tt 0811.0263}}.

\bibitem{Chowdhury:2010ct}
B.~D. Chowdhury and A.~Virmani, ``{Modave Lectures on Fuzzballs and Emission
  from the D1-D5 System},''
\href{http://arXiv.org/abs/1001.1444}{{\tt 1001.1444}}.

\bibitem{Bena:2013dka}
I.~Bena and N.~P. Warner, ``{Resolving the Structure of Black Holes:
  Philosophizing with a Hammer},''
\href{http://arXiv.org/abs/1311.4538}{{\tt 1311.4538}}.

\bibitem{Lunin:2001jy}
O.~Lunin and S.~D. Mathur, ``{AdS/CFT duality and the black hole information
  paradox},'' {\em Nucl. Phys.} {\bf B623} (2002) 342--394,
\href{http://arXiv.org/abs/hep-th/0109154}{{\tt hep-th/0109154}}.

\bibitem{deBoer:1998ip}
J.~de~Boer, ``{Six-dimensional supergravity on S**3 x AdS(3) and 2-D conformal
  field theory},'' {\em Nucl.Phys.} {\bf B548} (1999) 139--166,
\href{http://arXiv.org/abs/hep-th/9806104}{{\tt hep-th/9806104}}.

\bibitem{Bena:2015bea}
I.~Bena, S.~Giusto, R.~Russo, M.~Shigemori, and N.~P. Warner, ``{Habemus
  Superstratum! A constructive proof of the existence of superstrata},'' {\em
  JHEP} {\bf 1505} (2015) 110,
\href{http://arXiv.org/abs/1503.01463}{{\tt 1503.01463}}.

\bibitem{Bena:2016agb}
I.~Bena, E.~Martinec, D.~Turton, and N.~P. Warner, ``{Momentum Fractionation on
  Superstrata},''
\href{http://arXiv.org/abs/1601.05805}{{\tt 1601.05805}}.

\bibitem{Bena:2016ypk}
I.~Bena, S.~Giusto, E.~J. Martinec, R.~Russo, M.~Shigemori, D.~Turton, and
  N.~P. Warner, ``{Smooth horizonless geometries deep inside the black-hole
  regime},'' {\em Phys. Rev. Lett.} {\bf 117} (2016), no.~20, 201601,
\href{http://arXiv.org/abs/1607.03908}{{\tt 1607.03908}}.

\bibitem{Bena:2011dd}
I.~Bena, S.~Giusto, M.~Shigemori, and N.~P. Warner, ``{Supersymmetric Solutions
  in Six Dimensions: A Linear Structure},'' {\em JHEP} {\bf 1203} (2012) 084,
\href{http://arXiv.org/abs/1110.2781}{{\tt 1110.2781}}.

\bibitem{Mathur:2003hj}
S.~D. Mathur, A.~Saxena, and Y.~K. Srivastava, ``{Constructing `hair' for the
  three charge hole},'' {\em Nucl.Phys.} {\bf B680} (2004) 415--449,
\href{http://arXiv.org/abs/hep-th/0311092}{{\tt hep-th/0311092}}.

\bibitem{Mathur:2011gz}
S.~D. Mathur and D.~Turton, ``{Microstates at the boundary of AdS},'' {\em
  JHEP} {\bf 1205} (2012) 014,
\href{http://arXiv.org/abs/1112.6413}{{\tt 1112.6413}}.

\bibitem{Mathur:2012tj}
S.~D. Mathur and D.~Turton, ``{Momentum-carrying waves on D1-D5 microstate
  geometries},'' {\em Nucl.Phys.} {\bf B862} (2012) 764--780,
\href{http://arXiv.org/abs/1202.6421}{{\tt 1202.6421}}.

\bibitem{Shigemori:2013lta}
M.~Shigemori, ``{Perturbative 3-charge microstate geometries in six
  dimensions},'' {\em JHEP} {\bf 1310} (2013) 169,
\href{http://arXiv.org/abs/1307.3115}{{\tt 1307.3115}}.

\bibitem{Lunin:2012gp}
O.~Lunin, S.~D. Mathur, and D.~Turton, ``{Adding momentum to supersymmetric
  geometries},'' {\em Nucl.Phys.} {\bf B868} (2013) 383--415,
\href{http://arXiv.org/abs/1208.1770}{{\tt 1208.1770}}.

\bibitem{Giusto:2013bda}
S.~Giusto and R.~Russo, ``{Superdescendants of the D1D5 CFT and their dual
  3-charge geometries},'' {\em JHEP} {\bf 1403} (2014) 007,
\href{http://arXiv.org/abs/1311.5536}{{\tt 1311.5536}}.

\bibitem{Jejjala:2005yu}
V.~Jejjala, O.~Madden, S.~F. Ross, and G.~Titchener, ``{Non-supersymmetric
  smooth geometries and D1-D5-P bound states},'' {\em Phys. Rev.} {\bf D71}
  (2005) 124030,
\href{http://arXiv.org/abs/hep-th/0504181}{{\tt hep-th/0504181}}.

\bibitem{Giusto:2004ip}
S.~Giusto, S.~D. Mathur, and A.~Saxena, ``{3-charge geometries and their CFT
  duals},'' {\em Nucl. Phys.} {\bf B710} (2005) 425--463,
\href{http://arXiv.org/abs/hep-th/0406103}{{\tt hep-th/0406103}}.

\bibitem{Chakrabarty:2015foa}
B.~Chakrabarty, D.~Turton, and A.~Virmani, ``{Holographic description of
  non-supersymmetric orbifolded D1-D5-P solutions},'' {\em JHEP} {\bf 11}
  (2015) 063,
\href{http://arXiv.org/abs/1508.01231}{{\tt 1508.01231}}.

\bibitem{Mathur:2013nja}
S.~D. Mathur and D.~Turton, ``{Oscillating supertubes and neutral rotating
  black hole microstates},'' {\em JHEP} {\bf 04} (2014) 072,
\href{http://arXiv.org/abs/1310.1354}{{\tt 1310.1354}}.

\bibitem{Bossard:2014yta}
G.~Bossard and S.~Katmadas, ``{A bubbling bolt},'' {\em JHEP} {\bf 07} (2014)
  118,
\href{http://arXiv.org/abs/1405.4325}{{\tt 1405.4325}}.

\bibitem{Bossard:2014ola}
G.~Bossard and S.~Katmadas, ``{Floating JMaRT},'' {\em JHEP} {\bf 04} (2015)
  067,
\href{http://arXiv.org/abs/1412.5217}{{\tt 1412.5217}}.

\bibitem{Bena:2015drs}
I.~Bena, G.~Bossard, S.~Katmadas, and D.~Turton, ``{Non-BPS multi-bubble
  microstate geometries},'' {\em JHEP} {\bf 02} (2016) 073,
\href{http://arXiv.org/abs/1511.03669}{{\tt 1511.03669}}.

\bibitem{Bena:2016dbw}
I.~Bena, G.~Bossard, S.~Katmadas, and D.~Turton, ``{Bolting Multicenter
  Solutions},'' {\em JHEP} {\bf 01} (2017) 127,
\href{http://arXiv.org/abs/1611.03500}{{\tt 1611.03500}}.

\bibitem{Giusto:2013rxa}
S.~Giusto, L.~Martucci, M.~Petrini, and R.~Russo, ``{6D microstate geometries
  from 10D structures},'' {\em Nucl.Phys.} {\bf B876} (2013) 509--555,
\href{http://arXiv.org/abs/1306.1745}{{\tt 1306.1745}}.

\bibitem{Avery:2010qw}
S.~G. Avery, ``{Using the D1D5 CFT to Understand Black Holes},''
\href{http://arXiv.org/abs/1012.0072}{{\tt 1012.0072}}.

\bibitem{Skenderis:2006ah}
K.~Skenderis and M.~Taylor, ``{Fuzzball solutions and D1-D5 microstates},''
  {\em Phys.Rev.Lett.} {\bf 98} (2007) 071601,
\href{http://arXiv.org/abs/hep-th/0609154}{{\tt hep-th/0609154}}.

\bibitem{Roy:2016zzv}
P.~Roy, Y.~K. Srivastava, and A.~Virmani, ``{Hair on non-extremal D1-D5 bound
  states},'' {\em JHEP} {\bf 09} (2016) 145,
\href{http://arXiv.org/abs/1607.05405}{{\tt 1607.05405}}.

\bibitem{Balasubramanian:2000rt}
V.~Balasubramanian, J.~de~Boer, E.~Keski-Vakkuri, and S.~F. Ross,
  ``{Supersymmetric conical defects: Towards a string theoretic description of
  black hole formation},'' {\em Phys. Rev.} {\bf D64} (2001) 064011,
\href{http://arXiv.org/abs/hep-th/0011217}{{\tt hep-th/0011217}}.

\bibitem{Maldacena:2000dr}
J.~M. Maldacena and L.~Maoz, ``{Desingularization by rotation},'' {\em JHEP}
  {\bf 0212} (2002) 055,
\href{http://arXiv.org/abs/hep-th/0012025}{{\tt hep-th/0012025}}.

\end{thebibliography}

\end{document}